JPL Publication 18-1

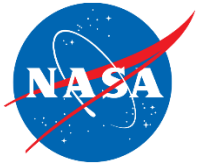

# Technical Note: Asteroid Detection Demonstration from SkySat-3* B612 Data using Synthetic Tracking

*A satellite owned/operated by Planet Labs, Inc.

C. Zhai (Jet Propulsion Laboratory), M. Shao (Jet Propulsion Laboratory), S. Lai (Planet), P. Boerner (Planet), J. Dyer (Google), E. Lu (B612 Asteroid Institute), H. Reitsema (B612 Asteroid Institute), M. Buie (B612 Asteroid Institute and Southwest Research Institute)

This research was funded by the B612 Asteroid Institute and carried out at the Jet Propulsion Laboratory, California Institute of Technology, an institution under contract with the National Aeronautics and Space Administration.

**Abstract**

We report results from analyzing the B612 asteroid observation data taken by the sCMOS cameras on board of Planet SkySat-3 using the synthetic tracking technique. The analysis demonstrates the expected sensitivity improvement in the signal-to-noise ratio of the asteroids from properly stacking up the the short exposure images in post-proessing.

# CONTENTS

# 1    BACKGROUND

The B612 Asteroid Institute funded a Caltech research project relating to the preparation of a feasibility study for a SmallSat NEO survey mission using synthetic tracking [1]. The research was of mutual interest and benefit to both organizations, and helped further research objectives with potential resulting benefits.

The B612 Asteroid Institute engaged leadership at Terra Bella/Google to partner on a demonstration project in support of the Institute's mission to protect the planet from asteroid impacts using the synthetic tracking technique developed through its prior research project with Caltech.

SkySat-3 is an Earth-observing satellite launched in 2016 by Terra Bella Technologies (formerly Skybox Imaging, Inc.). Skybox was acquired by Google in ~2015. The B612 Asteroid Institute worked with Terra Bella/Google to turn one of the Skybox satellites around to look skyward instead of downward at Earth to test the technique of Synthetic Tracking [2][3] from space to detect Near Earth Asteroids. From January to April 2017, six half-orbits of data were obtained from SkySat-3. In early 2017, Google sold their Terra Bella subsidiary to Planet Labs, Inc. Planet now owns and operates a constellation of 13 of these satellites, the 1st of which to have propulsion was SkySat-3.

This report describes the analysis of this data and demonstrates the technique of synthetic tracking for distinguishing dim moving asteroids with an increase in sensitivity consistent with theory.

# 2  SKYSAT TELESCOPE CAMERA

SkySat-3 has a 35cm diameter RC telescope and three Fairchild/BAE sCMOS detectors. The focal length was chosen so that one pixel was roughly the diffraction limit of the 35cm telescope.

- Telescope diameter: 350mm
- Focal length: 3.6m (f/10.3)
- Pixel size: 6.5um, 0.372 arcsec in sky
- Detector array size: (1/2 of chip was panchromatic, 450-900nm passband) 1080*2560 pixels
- 3 sCMOS chips in the focal plane
- Operating mode: video at 50 frames/s (20msec exposure) in global shutter mode 5e read noise
- Detector gain: 1.8e/dn

When observing the Earth, the camera can produce images with ~0.9 m resolution at 500km altitude, implying the optical figure and alignment/focus of the optics produced a diffraction limited image.

# 3 SPACECRAFT ACS, ORBIT, AND DATA COLLECT SCENARIO

Pointing stability is an important parameter for a space-based telescope searching for NEOs[1]. Terra Bella provided on-orbit data to characterize a pointing jitter of the ACS system on SkySat-3. The plot below shows the pointing jitter as measured by the science camera taking frames at 50hz on a bright star. The science camera data was not used to provide feedback to the reaction wheels.

The Y axis of the plot is motion in arcsec. The RMS pointing jitter was about 1 arcsec, although the fluctuations are certainly not a gaussian random process. Still 1 arcsec RMS is quite a bit better than most CubeSat type ACS systems[4]. But as one can see in the graph, the peak-to-peak motion can be as large as 8 arcsec.

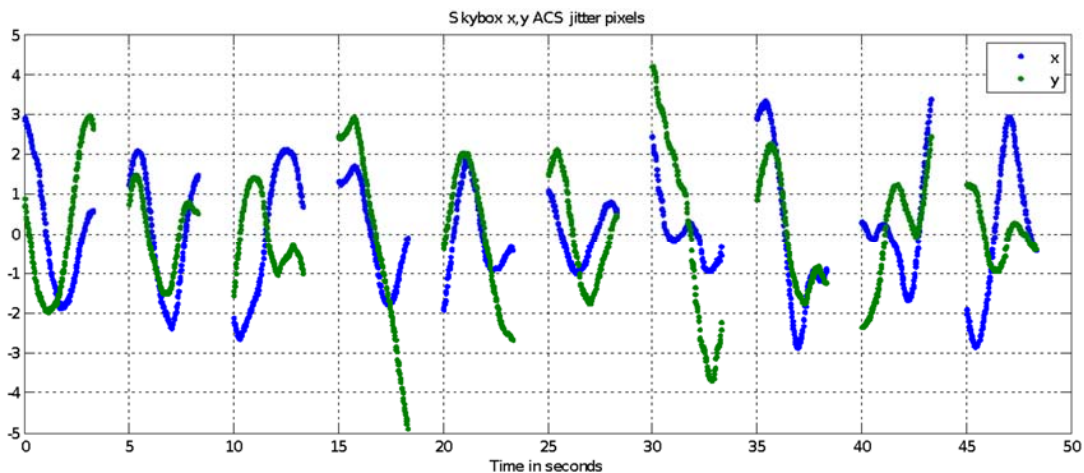

**Figure 1. Telescope pointing as function of time. Y-axis units is arcsec.**

SkySat-3 is in a polar sun synchronous orbit, passing above the Equator at ~10 a.m. This orbit puts the satellite in the shadow of the Earth over roughly 50% of its orbit. The asteroid observations were made when the satellite was in Earth's shadow, when its Earth observing mission could not be conducted.

# 4    SKYSAT-3 OBSERVATIONS

SkySat-3 made 6 (½) orbit observing runs between January 25 and April 5 of 2017. The camera was operated in its normal (Earth observing) mode taking data at 50Hz. Because of raw (uncompressed) data recording and downlink limitations, only 3.3 seconds of data could be recorded in RAM at one time; that is 166 frames of video. The three sCMOS detectors and their electronics and memory could hold 3.3 seconds of uncompressed data each.

The three sCMOS detectors have 2560*2160 pixels each. ½ of the detector was covered by a multispectral color filter. The 4 different color filters make "color" images of the Earth possible by scanning the scene through each of the color filters. For our NEO observations, a known asteroid was placed in the panchromatic (450-900nm filter) ½ of the detector with 2560*1080 pixels.

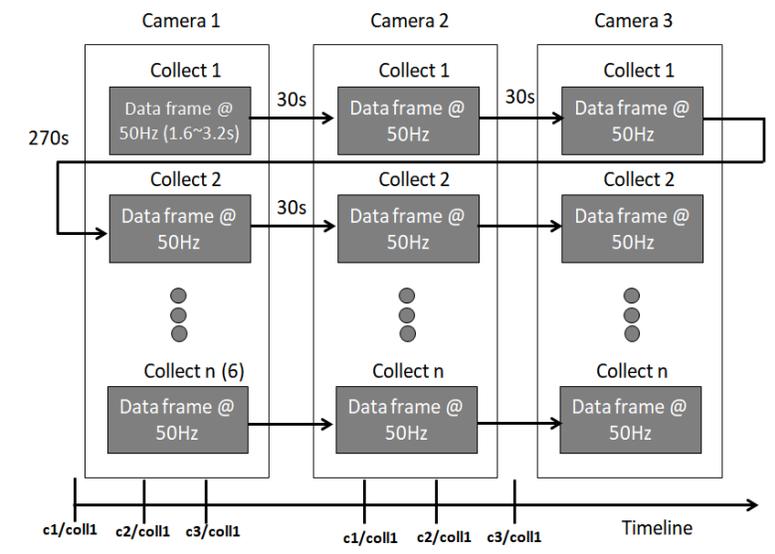

**Figure 2. Observation Cadence.**

Data was collected by pointing at the predicted position of a known asteroid, first in detector #1, collecting data for 3.3 seconds and then moving the telescope (~30s) to place the same asteroid in detector #2, taking another 3.3 seconds of data, and so on. Only one asteroid was observed at each pass. The data collection is illustrated in the figure above.

The data set analyzed here for one asteroid observation would consist of a total of 50 seconds of data, taken in 3.3 second chunks spread over ~23 minutes. Within each 3.3 seconds of video, the pointing of the telescope would jitter on average ~1 arcsec, but could on occasion, move as much as 7 arcsec, or 19 pixels. With a single 20 msec exposure the image smearing due to ACS jitter would always be less than 0.13 pixels and usually less than 0.05 pixels.

When a bright star was in the focal plane, the image jitter from the attitude control system could be removed to +/- 0.5 pixel.

# 5 DATA PROCESSING

The above data collection procedure for SkySat-3 was very different from our previously demonstrated ground based synthetic tracking observations, which typically consist of 120 frames of video taken at 1 Hz. In our ground-based observations, synthetic tracking is applied to a contiguous video sequence, with no gaps. The multi-vector shift and add algorithm running on a GPU was designed to accept evenly-sampled video data[3]. The analysis of SkySat-3 data is a one-time effort.

The main steps of data processing of SkySat-3 consists 1) estimate sky background and subtract it from each data frame and remove signals due to cosmic ray events and bad pixels; 2) re-register the 50Hz frames to compensate the pointing jitters so that stars are aligned within each data collect; 3) integrate frames from all the data collects.

We will illustrate the data processing via the analysis of two data sets, one for the main belt asteroid, Erynia, and the other for the NEO 1998YP11.

## 5.1 ANALYSIS OF ERYNIA DATA

### 5.1.1 *BIAS FRAME ESTIMATION*

We first estimate the bias and subtract it from all the frames. A first-order correction of temperature-dependent gain and bias is performed on-board using a look-up table. The dark current in 20 msec was negligible, the zodi background from space >22 mag/arcsec$^2$ was also negligible in 20msec. Dark frames were not available and the engineers at Terra Bella suggested using a median filter to remove residual bias. The median filter works because of the spacecraft jitter. The detector bias and sky background is estimated by looking at every pixel in time over the whole data set. Because the pointing jitter is >> FWHM of a stellar image, every pixel (even the ones near stars) spends most of the time looking at space. If we take 1 pixel versus time and plot a histogram of DN of that pixel, the peak of that distribution represents that pixel looking at a dark sky. The estimated bias frame of the detector is shown in Figure 3 below.

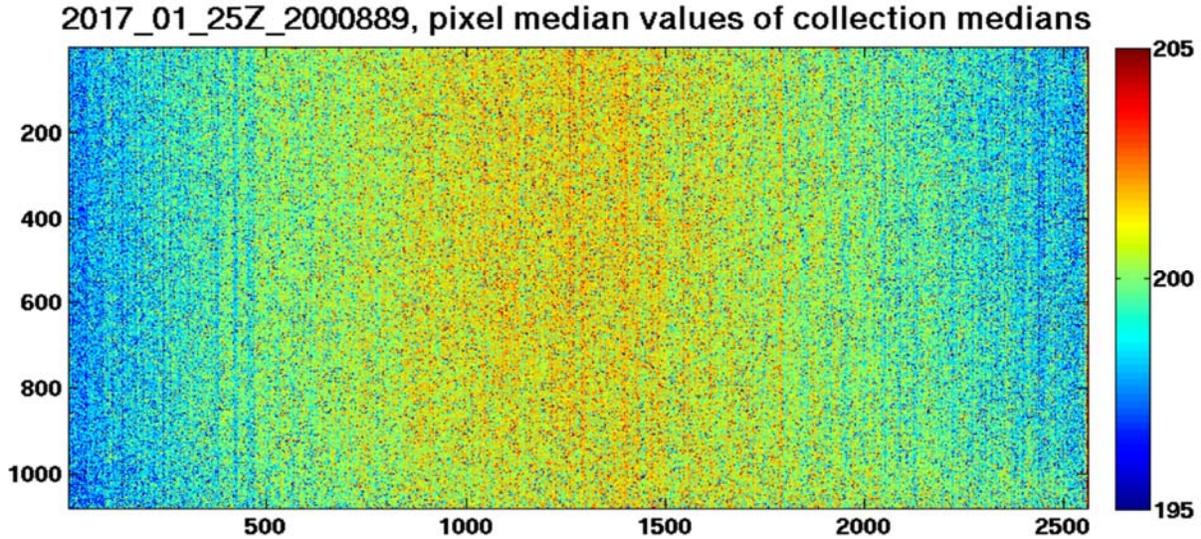

**Figure 3. Estimated sky background using multiple data collects. For each pixel, we first estimate the median value of each data collect and then take the median of all the data collect median values. The false color scale is in DN. The vertical streaks in the bias frame are typical for sCMOS detectors.**

## 5.1.2 *FRAME RE-REGISTRATION*

For Erynia, (apparent magnitude 14.5 mag), there was a very bright 12.8 mag star in the FOV, which was clearly detected in every 20 msec frame. This allowed the jitter of the spacecraft ACS to be removed +/- 0.5 pixels.

While the brightest 1~2 stars were unambiguous, we wanted to clearly see a half dozen stars to provide an astrometric reference frame. Unfortunately, the faintest of these looked the same as "hot" pixels in the detector as shown in Figure 4.

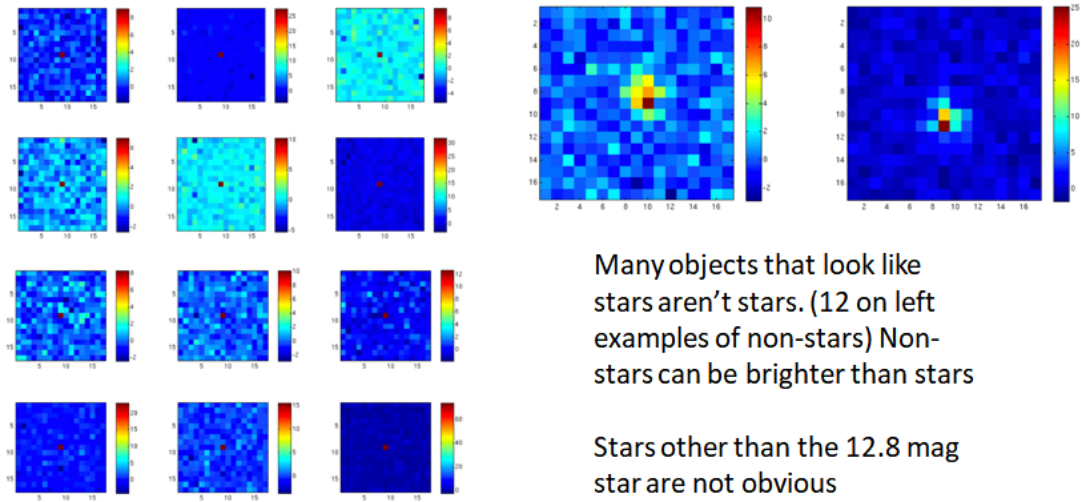

Many objects that look like stars aren't stars. (12 on left examples of non-stars) Non-stars can be brighter than stars

Stars other than the 12.8 mag star are not obvious

**Figure 4. Sample sub-frame images (50Hz) of signals above noise level. At left, 12 signals are due to either cosmic ray or bad pixels and the two images at right side are star images in a single 50Hz frame.**

Many hot pixels are brighter than stars. Again one can make use of the pointing jitter of the spacecraft. Hot pixels are fixed to the focal plane, not fixed to the sky. A 16th mag star would, on average, have ~8 detected photons per 20msec frame, where the read noise was 5 electrons. This sCMOS detector has a 5e average read noise. But 5e is the median read noise. ½ the pixels on the chip has > 5e and the other ½ < 5e read noise. Typically in sCMOS sensors a non-trivial fraction (1e-3) of pixels can have read noise more than 2~3 times the "average" read noise. These high noise pixels can be identified if the detector is carefully characterized in the lab prior to launch of the spacecraft. But they are hard to identify. We masked out the pixels whose temporal variations are consistently above 5 times of the average pixel temporal variations in all the data collects.

### 5.1.3 *SYNTHETIC IMAGE OF ERYNIA*

After we re-register the frames, we integrate the aligned 83 frames of each data collect by taking a simple average of these frames to track the stars. This is justified because Erynia is a main belt asteroid at ~1.3 AU from the Earth at this time of observation (2.3 AU from the Sun) according to JPL Horizon System[5], its motion over 1.6seconds is much smaller than a pixel.

With four images from integrating four data collects separated by about 4.5 minutes, we can further stack up them using synthetic tracking assuming that it motion over the four data collects (~14 minutes) is close to linear; this is a good approximation because Erynia was far enough away that the motion of the spacecraft around the earth (orbital parallax) was relatively small. The magnitude of the RA, Dec of Erynia change due to the spacecraft's orbit could be as large as 11 arcsec, but over the 14 minute time span of the data collection, the non-linear component of motion was only on the order of 1 pixel. However, the apparent angular velocity of Erynia as seen by SkySat-3 was markedly different than the angular velocity as seen by an observer on the ground.

We searched over a 100 x 100 velocity grid with spacing ~ 1 pixel per integration time (14 minutes) to integrate the 4 frames via the GPU-aided multi-vector shift/add. With 10,000 velocity vectors, we have 10,000 synthetic images and GPU reported 13 signals with peaks above 15x temporal noise. This is higher than our typical 7x threshold because the spatial noise level is higher than the temporal noise, which we will discuss later.

Clustering the 13 signals yields only one strongest signal, with both estimated astrometry and rate matching that of Erynia. The left image in Figure 5 displays the integrated signal of Erynia assuming linear motion over the ~14 minute collection period.

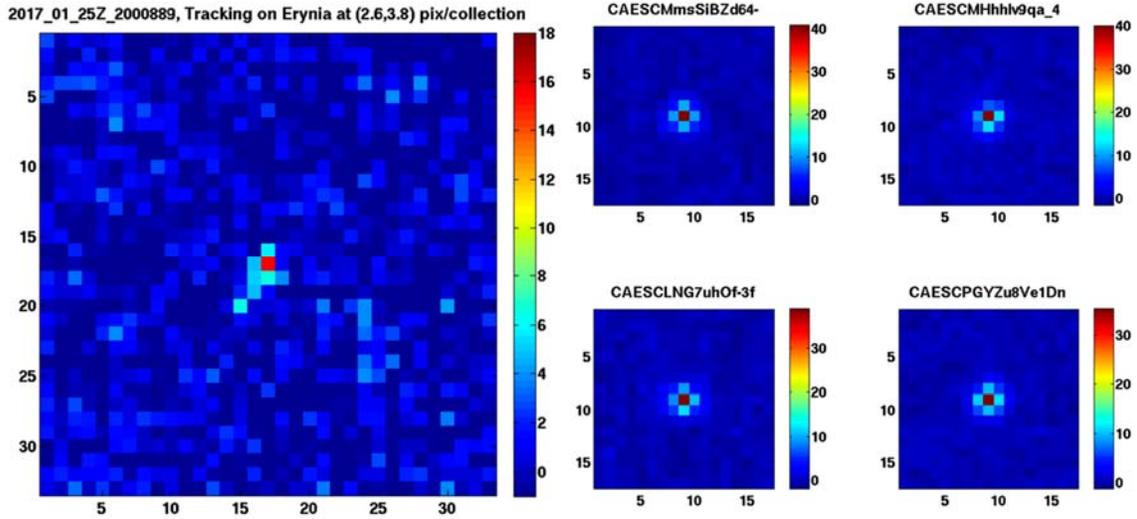

**Figure 5. Images from stacking up all the Erynia images (left) and images from stacking up images in each data collect.**

The PSF is not as sharp as the image of a stacked star shown on the right due to a small amount of non-linear motion from orbital parallax over the 14 minute time collection. Erynia was bright enough that it was detectable without using all the data. It was easily detected using just 1.6 seconds of data (there are only 83 frames at 50Hz for each data collect of the Erynia's observation). Its motion over ~14 minutes is shown in Figure 6.

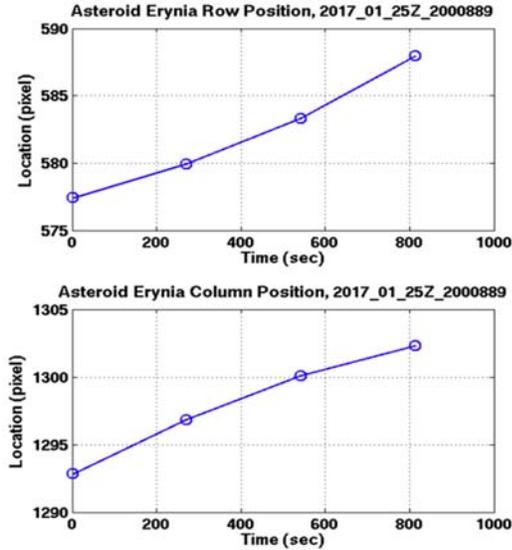

**Figure 6. Asteroid Erynia positions in camera frame as function of time.**

The slight non-linear motion is visible in the plots. Each dot represented 1.6 seconds of 50 Hz video. 4 data points over 800 seconds were recorded. The images of those 4 points are shown in Figure 7. The left image is a composite of the 4 data collects. The center one shows where Erynia was with golden circles. The image on the right are the 4 images of Erynia co-added using the known non-linear motion because we the asteroid was detectable using 1.6 seconds of data.

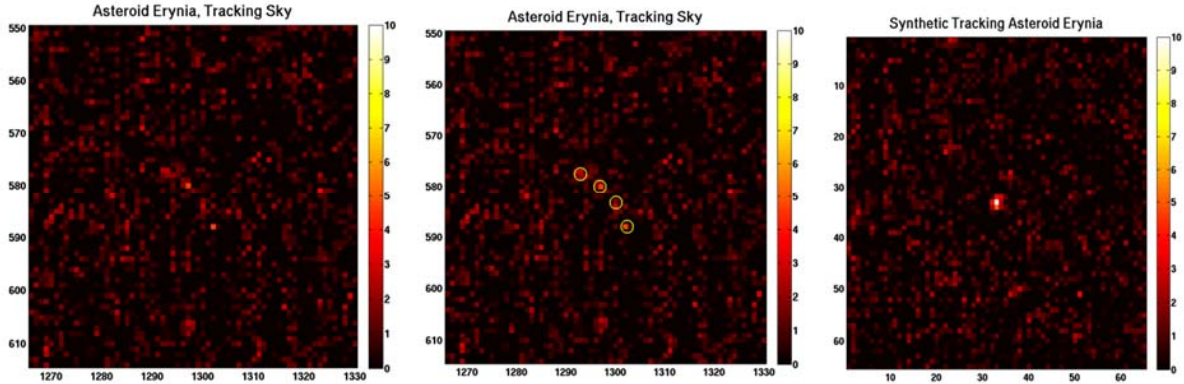

**Figure 7. Synthetic tracking images from tracking the sidereal (left and mid, without and with Erynia circled), and tracking Erynia (right).**

A typical 4K*4K CCD has 4 readout amplifiers. Each read amp is used to read the charge in ~4 million pixels. The charge transfer process in a CCD is near perfect and the read noise is entirely in the read amp. An sCMOS detector has a different read amp for every pixel. In a CCD, every pixel has a slightly different QE, but adjacent pixels have the same gain and the same noise. In sCMOS, all of these quantities are different from pixel to pixel. Read noise in CCD and sCMOS is the variation in the output when 1 pixel is read repeatedly. This is also called temporal noise. Spatial noise is the pixel-to-pixel variation in a single frame after bias subtraction. When detecting a NEO in an image, spatial noise is what is important, not temporal noise. In most science grade CCDs, spatial noise is equal to temporal noise. In general this is not the case for sCMOS. In CCD detection of NEOs, one attempts to detect the NEO in a single frame. If the read noise (temporal) noise is 3e, we need to understand temporal noise at a level better than 3e. But in synthetic tracking, we can average 100 or in the case of SkySat-3 data, 2490 frames of data and we expect/hope that by averaging 2490 frames the SNR improves by ~50X, from 5e to 0.1e. For our ground based observations, we've developed observation and detector calibration approaches that let us produce synthetic images (co-added frames) with spatial noise just 10% larger than temporal noise. However for the SkySat data set we find the spatial noise is 2~3 times larger than the temporal noise.

## 5.2 NEO 1998YP11

For 1998YP11, there wasn't a bright star in the FOV that was visible in every 20 msec video frame; the brightest star is of ~14th mag. However if we look at the spacecraft attitude jitter plots we notice that the pointing drift is linear for 0.2 seconds almost all the time. That means we can use synthetic tracking (10 frames at a time) to increase the SNR of the stars. Sqrt(10) buys us slightly more than a stellar magnitude to find stars, which then can be detected at reasonable SNR and re-registered every 10 frames.

The apparent magnitude of 1998YP11 is 16.1mag and cannot be detected in a single 3.3 second video sequence or 3 successive collects (~10 sec) integration. Figure 8 shows 5 collects each representing 10 sec of observation. (The 6th data collect was not used due to existence of extra noise possibly from cosmic ray). As mentioned before, the spatial noise does not look like white noise and in fact is 2~3 times higher than the temporal noise.

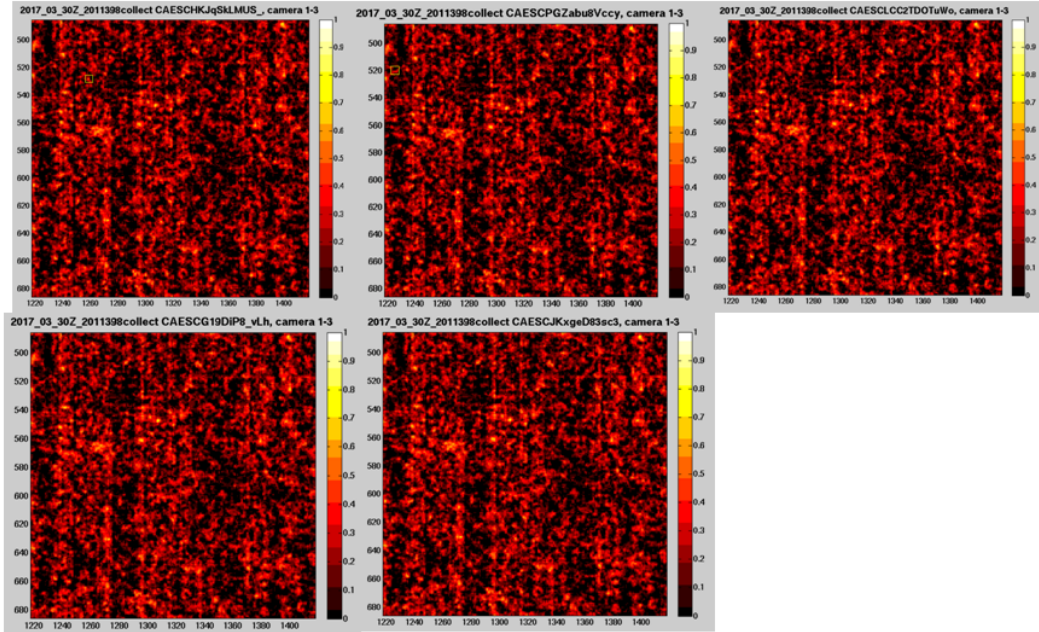

**Figure 8. Five synthetic tracking images from integrating five data collects. Each data collect includes data from the three cameras.**

Synthetic tracking as currently practiced, takes a 3D data cube (N images in a video sequence) and searches for a moving object in 4D space, x, y, Vx, and Vy. For this data set, that would not work because this object is only 0.3 AU away, not 1.3AU, and the orbital parallax effect from the satellite, which over a full orbit would be 30arcsec peak-to-peak. It is possible to extend synthetic tracking to include an extra $5^{th}$ dimension, the distance to the NEO. With the known orbital ephemeris of SkySat-3, a 5-D search would be able to find a NEO at 0.3 AU from a satellite in LEO. The apparent position and motion of 1998YP11 using the known satellite ephemeris (upper left) and NEO ephemeris (upper right) is shown in Figure 9 below.

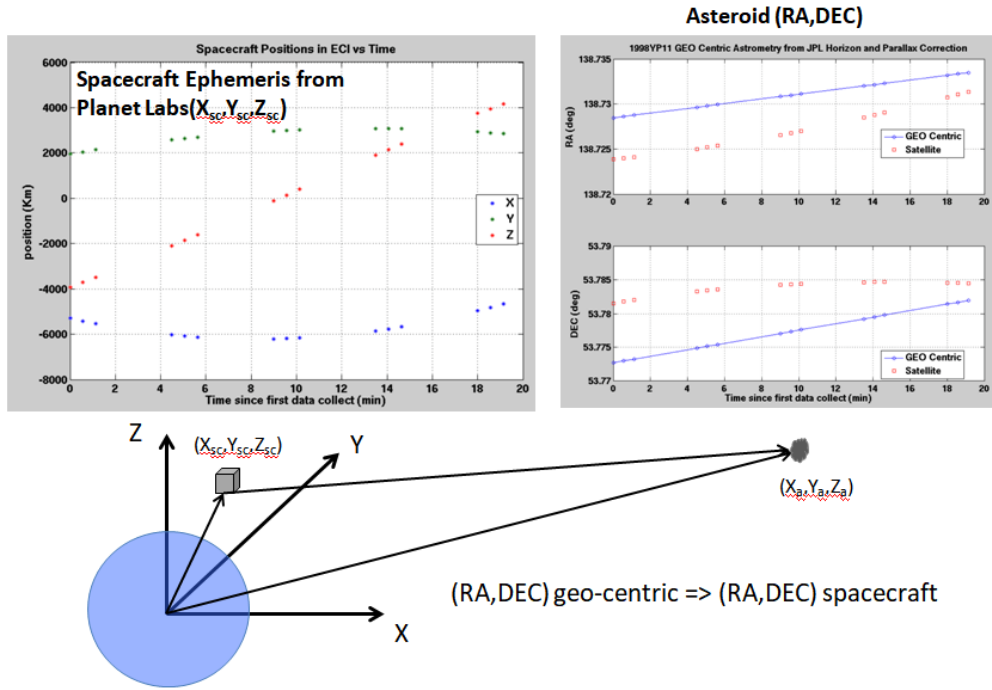

**Figure 9. Ephemeris of SkySat3 from Planet (upper left), asteroid 1998 YP11 sky position versus time as observed from the GEO centric location (blue) and the SkySat3 (red), and a schematic of observation geometry.**

When all 2490 frames are properly stacked, one gets the image in Figure 10 at the right side. The left image below, with an improved SNR, is obtained by convolving the image on the right with a kernel, which is the PSF of the brightest star in the field from stacking up all the frames. The false color image is in units of 1 sigma spatial noise (over the 1K*2.5K pixel image).

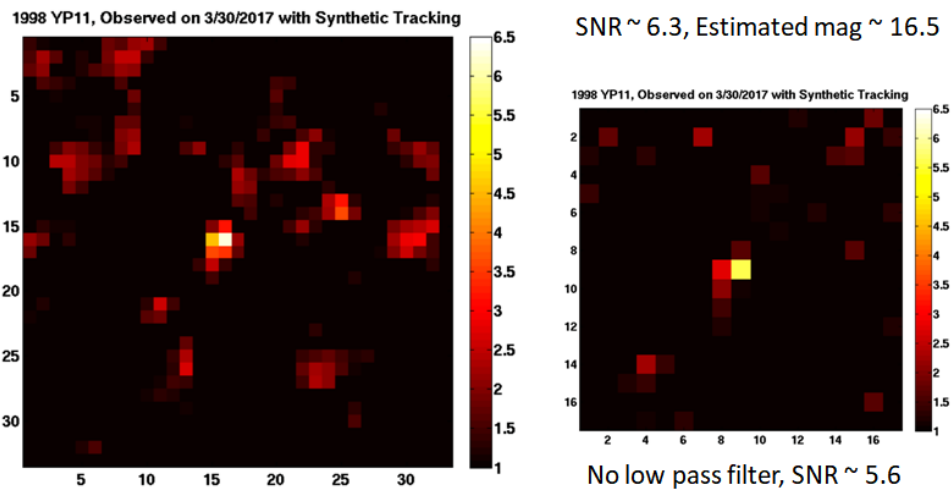

**Figure 10. Synthetic tracking image from stacking up all the images (right) and its convolution with a kernel proportional to the PSF of the brightest star to improve SNR (right).**

# 6 DISCUSSIONS AND FUTURE PERSPECTIVE

## 6.1 IDEAL SNR FROM A SKYSAT TYPE CAMERA

The actual SNR for 1998YP11 was SNR~6.3. But the theoretical/ideal SNR should have been much higher. If we assume that the frame/frame star stacking error would only widen the PSF to 2*2 pixels, and the temporal noise of 5e/frame was also the spatial noise, then the SNR for stacking 2490 frames should have been SNR~28. While we can't say exactly, very likely spatial noise was 2~3 times worse than temporal noise and stacking the stars to remove spacecraft jitter resulted in a PSF slightly worse than 2*2 pixels.

## 6.2 MORE OPTIMIZED OPERATING PARAMETERS FOR SUCH A CAMERA

The SkySat camera was not operated in a way that was optimal for NEO detection. The limiting factor was the very short 20 msec exposure time, a consequence of the camera firmware having being designed for Earth observation. While read noise was 5e equal to a background flux of 25 photons/pixel/frame, the Zodi background was only 0.01 photons/frame/pixel. If one were able to rearrange the camera hardware and operating mode, a 35cm telescope with 1 arcsec pixels running with a rolling shutter (1.5e vs 5e read noise) with 2~5 second exposure, would have a limiting magnitude of 21.7 mag at SNR=7, in a 50 second data cube. These modifications are possible on the SkySat platform but would require some changes to firmware which was out of scope for this exercise. 50 seconds is also short enough that virtually all NEOs would move in a straight line, even with orbital parallax. If the sensor was a 4K sensor, which is now available off the shelf, the FOV would be ~1.3 sqdeg (versus 0.03 sqdeg).

There were several reasons for a tentative choice of 1 arcsec pixels. The larger angular extent of this pixel (versus 0.37 arcsec) along with the longer single frame exposure time > 2 sec versus 0.02 sec means we are now safely in the zodi background noise limited regime instead of detector noise limited operation. The other was that the 1 sigma RMS jitter of SkySat-3 was 1 arcsec. With a less capable ACS system one would likely want larger pixels.

## 6.3 FUTURE DIRECTIONS IN ASTEROID SEARCHES

Recently Pan-STARRS announced the first discovery of an interstellar asteroid. This came after ~10 years of Pan-STARRS operation. Early estimates are that the discovery rate would only increase slightly in the near future, even with telescopes like LSST. The principal reason is that interstellar asteroids move very rapidly compared to normal NEOs, thus producing a streaked image with Pan-STARRS and LSST. Synthetic tracking would be able to increase the discovery rate by 10~30, even when using relatively small telescopes such as the 35cm aperture SkySat-3 telescope.

# ACKNOWLEDGEMENTS

The Asteroid Institute is a program of the B612 Foundation, and the funding is generously provided by the W.K. Bowes Jr. Foundation and Steve Jurvetson along with Founding Circle and Asteroid Circle members K. Algeri-Wong, B. Anders, G. Baehr, B. Burton, A. Carlson, D. Carlson, S. Cerf, V. Cerf, Y. Chapman, J. Chervenak, D. Corrigan, E. Corrigan, A. Denton, E. Dyson, A. Eustace, S.

Galitsky, The Gillikin Family, E. Gillum, L. Girand, Glaser Progress Foundation, D. Glasgow, J. Grimm, S. Grimm, G. Gruener, V. K. Hsu & Sons Foundation Ltd., J. Huang, J. D. Jameson, J. Jameson, M. Jonsson Family Foundation, D. Kaiser, S. Krausz, V. Lašas, J. Leszczenski, D. Liddle, S. Mak, G.McAdoo, S. McGregor, J. Mercer, M. Mullenweg, D. Murphy, P. Norvig, S. Pishevar, R. Quindlen, N. Ramsey, P. Rawls Family Fund, R. Rothrock, E. Sahakian, R. Schweickart, A. Slater, T. Trueman, F. B. Vaughn, R. C. Vaughn, B. Wheeler, Y. Wong, M. Wyndowe, and eight anonymous donors.

We thank Melissa Klose at JPL for her excellent editorial work, Lea Takigawa and David Lee for reviewing the manuscript. Their support is provided by NASA-JPL-Division 27.

## REFERENCES

[1] Michael Shao, Slava G. Turyshev, Sara Spangelo, Thomas Werne, and Chengxing Zhai, A constellation of SmallSats with synthetic tracking cameras to search for 90% of potentially hazardous near-Earth objects, A&A 603, A126, DOI: 10.1051/0004-6361/201629809 (2017).

[2] M. Shao, B. Nemati, C. Zhai, S. Turyshev, J. Sandhu, G. Hallinan, L. Harding, Finding Very Small Near-Earth Asteroids Using Synthetic Tracking, ApJ, 782, 1, (2014).

[3] C. Zhai, M. Shao, B. Nemati, T. Werne, H. Zhou, S. Turyshev, J. Sandhu, G. Hallinan, L. Harding, Detection of a faint fast-moving near-Earth asteroid using synthetic tracking technique, ApJ 792, 60 doi:10.1088/0004-637X/792/1/60, http://arxiv.org/abs/1403.4353 (2014).

[4] http://bluecanyontech.com/wp-content/uploads/2017/07/DataSheet_ADCS_08_F.pdf

[5] https://ssd.jpl.nasa.gov/?horizons_doc